\input{aipcheck}

\documentclass[
    ,final            
  ]
  {aipproc}

\layoutstyle{8x11single}

\begin{document}

\title{{Light $O^{++}$ Mesons: Scalargators in Florida}}


\author{M.R. Pennington}{
  address={Institute for Particle Physics Phenomenology, Physics Department, Durham University, 
        Durham DH1 3LE, U.K.}}
\begin{abstract}
{Light scalar mesons abound in hadron processes, like the alligators in the Florida Everglades. Moreover, scalars are intimately tied to the vacuum structure of QCD. They are the product of many decays. Consequently, a rich source of recent information about them has come from experiments producing heavy flavour mesons. Indeed, scalars will continue to dominate many of the processes to be studied at forthcoming facilities like BESIII in Beijing, FAIR at GSI Darmstadt and the GlueX experiment at JLab, making an understanding (or at least an excellent and theoretically consistent description) essential for the physics missions of these facilities.
} 
\end{abstract}
\maketitle

\section{Why hunt scalargators?}
Scalars are interesting critters, fleet of foot and hard to spot, yet they lie everywhere in almost every hadronic reaction. By emitting a $O^{++}$ meson, any heavy particle can decay, particularly if the scalar doesn't even change the flavour quantum numbers of the initial state. Consequently, the most ubiquitous scalars are those that have vacuum quantum numbers. Indeed, these are intimately tied to the nature of the QCD vacuum. We have long known that this vacuum is not empty. Over distance scales of the size of a hadron the interactions between gluons, quarks and antiquarks are so strong that they polarise the vacuum, forming condensates that populate the ground state. While an {\it up} or {\it down} quark propagates over short distances (like a hundredth of a fermi) as though it were almost massless with interactions that can (thanks to asymptotic freedom) be treated perturbatively, over longer distances of the order of a fermi, these quarks have to travel through a medium filled with these condensates. These slow their progress generating a ``constituent'' mass of some 350 MeV for the $u$ and $d$ quarks.
Such dynamical mass generation is of course only possible in a strong coupling field theory.  Which particular condensate of quark, antiquark and gluon controls this behaviour can not only be calculated in QCD, but also measured in experiments~\cite{swimming,eft09}.

The behaviour of the quark propagator, and in particular its mass function, is calculable as a function of momentum using the field equations of QCD, namely the Schwinger-Dyson equations. These computations can be performed both in the continuum~\cite{bowman} and with a lattice regulator~\cite{maris-roberts}. Where these calculations overlap, typically when $m_q(M_Z) \ge 25$ MeV, they agree. However, it is only in the continuum that one can consider realistically light quarks, since these do not fit on a limited size lattice.   The momentum dependence of the quark mass function with a current mass $\sim 3$ MeV, like the real {\it up/down} average, is very close to that for massless quarks. However, it is in the massless limit that we can use the Operator Product Expansion and learn that the behaviour of the mass function is in fact controlled by the $\langle \,{\overline q} q \, \rangle$ condensate with a value of $\sim - (240\, {\mathrm{MeV}})^3$, where the scale is set by $\Lambda_{QCD}$~\cite{swimming,williams,maris-roberts}.
 This determines the physics of low energy meson interactions. 

If the {\it up} and {\it down} quarks were really massless, QCD has a chiral symmetry: the quarks spinning left-handedly decouple from those spinning right.  If this symmetry were imparted to the hadron world then scalars and pseudoscalars, vectors and axial-vectors, would be degenerate in mass with closely related interactions. Very obviously this is not realised in nature. Indeed, pions are  much lighter than any other hadron. With dynamics controlled by the square of the mass, pions are 25 times lighter than a typical ${\overline q}q$ meson, like the $\rho$.
 The chiral symmetry of QCD at the quark level is dynamically broken in the world of hadrons.  Scalars and pseudoscalars are quite different. In a world of massless quarks, pions would be massless. However, scalars, like all other hadrons,  feel the \lq\lq constituent'' mass of quarks, generated, as we have seen, largely by the ${\overline q}q$ condensate. 

Pions, being the Goldstone bosons of chiral symmetry breaking, know about the nature of the condensates that do this breaking. Their interactions at low energy reflect these. Moreover, being the lightest of all hadrons, their interactions are universal (independent of the way they are produced).    
Access to low energy $\pi\pi$ final state interactions is provided by semileptonic decays (which as we will see is a recurring theme of this talk), in particular in $K_{e4}$ decay: $K \to e \nu_e (\pi\pi)$. By studying this decay distribution as a function of its  5 kinematic invariants, we can learn about the energy dependence of the $\pi\pi$ $S-P$ phase difference extracted from the BNL E865 experiment~\cite{bnl-e865} and with even greater precision from CERN NA48/2~\cite{na48}. When combined with our knowledge of chiral perturbation theory~\cite{gasser}, this tells us that the vacuum condensate is indeed $\langle \,{\overline q} q \, \rangle\,\simeq\, - (240\, {\mathrm{MeV}})^3$, just as calculated in strong coupling QCD. The vacuum breaks the chiral symmetry. The explicit breaking produced by the small current mass of the {\it up}/{\it down} quark ensures the physical pion is light with a mass$^2$ of just $0.02$ GeV$^2$.

\section{Scalars sighted}

In the simplest model of symmetry breaking, like that proposed by Nambu~\cite{nambu}, this pion has a single scalar partner, called the $\sigma$. This is naturally identified with the carrier of the isoscalar nuclear force. But is there just one scalar that plays the role of partner to the pion? Whilst $I=J=0$ $\pi\pi$ scattering from 600 to 1800 MeV was determined by the classic CERN-Munich experiment~\cite{ochs,hyams,grayer} decades ago, it is the information from semileptonic decays~\cite{bnl-e865,na48} that fixes its near threshold behaviour. When this is combined with the constraints on the nearby left hand cut imposed  by crossing symmetry, Caprini {\it et al.}~\cite{caprini-pole} have been able to locate the corresponding pole of the $S$-matrix at $E =\, 441 -i 272$ MeV. While there has been debate~\cite{zzs,peleaz} about whether the uncertainties in this position are $\pm 15$ MeV or $\pm 25$ MeV, we know this is near where the pole lies. It is this position that translates from one process to another. The most important thing to note about the $\sigma$ is how very close in terms of $s=E^2$ it sits to the threshold for the $\pi\pi$ channel, to which it strongly couples and quickly decays. 

\vspace{2mm}
\begin{figure}[h]
\includegraphics[width=0.64\textwidth]{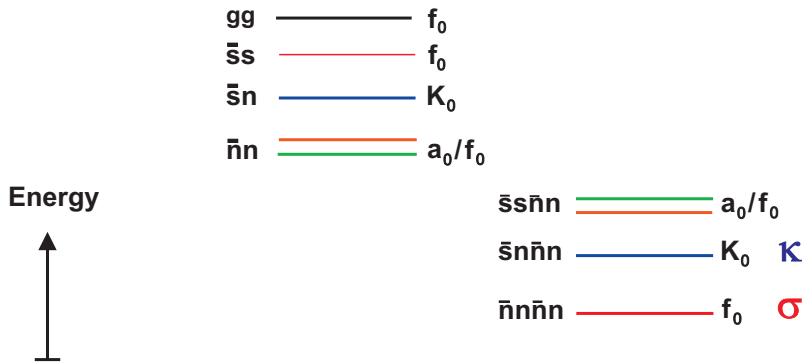}
\vspace{6mm}
\caption{The spectrum of scalar states given by a simple {\it ideal} nonet of ${\overline {q}}q$ and of ${\overline{qq}qq}$ mesons, where $n = u,d$, indicating which would be identified with the isoscalar $\sigma$ and its isodoublet partner, the $\kappa$, if the observed hadrons had these presumed {\it orthogonal} compositions. Where a glueball state, $gg$, might lie is also indicated. }
\end{figure}
\vspace{1mm}

However, this $\sigma$ is just one of a series of isoscalar states~\cite{pdg}: $f_0(980)$, $f_0(1370)$ (if it exists -- see ~\cite{bugg1370,ochs1370}), $f_0(1510)$, $f_0(1720)$, $\cdots$. These are accompanied by isotriplets $a_0(980)$ and $a_0(1430)$,  and isodoublets $K^*_0(1430)$ and the low mass $\kappa$. Clearly far more states than can fit into one ${\overline q}q$ multiplet. Indeed, they might form two nonets, with possibly one additional state left over to be a glueball candidate. But which is which?

Long ago Jaffe~\cite{jaffe-old} noted that four quark states ${\overline {qq}}qq$ might well exist. More recent work~\cite{jaffe-wilczek} has discussed such mesons in terms of diquark-antidiquark systems.  Two quarks in a ${\overline 3}$ of colour will bind if they have different flavours, in keeping with Pauli exclusion, to form a scalar diquark. A triplet of scalar diquarks $[ud]$, $[ds]$ and $[su]$ then attract anti-diquarks to form colour singlets in the shape of a tetraquark nonet. For $O^{++}$ quantum numbers, Jaffe noted that this multiplet would be at lower mass than that of just a quark and antiquark with $L_{{\overline q}q}=S_{{\overline q}q}=1$, as depicted in Fig.~1.
This picture seems to fit the experimental information, not just in terms of counting, but provides an explanation of how two well-known states, the $f_0(980)$ and $a_0(980)$, can be degenerate in mass and both couple strongly to ${\overline K}K$: something difficult to understand for an ${\overline n}n$ isotriplet (where $n\,=\,u,\,d$)  and a largely ${\overline s}s$ state. In contrast, for a $[{\overline{sn}}][sn]$ system this is totally natural. This explanation of the low mass scalars thus seems very plausible and has been much discussed in the literature~\cite{schechter,maiani}.
\begin{figure}[h]
\vspace{2.5mm}
\includegraphics[width=0.8\textwidth]{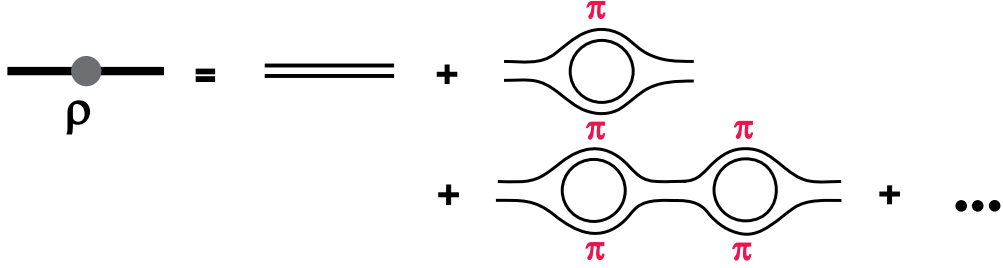}
\vspace{2mm}
\caption{The propagator of the $\rho$-meson expressed in terms of quark line graphs. At lowest order it is assumed to be a ${\overline q}q$ meson, which decays at higher order by coupling to pion pairs.}
\vspace{-2mm}
\end{figure}

However, one should perhaps probe a little closer into the relation between underlying quark model states and the hadrons we observe. The paradigm for the structure of  a ${\overline q}q$ multiplet is the nonet of light vector states: $\rho$, $K^*$, $\omega$ and $\phi$. These form a beautiful  ideally mixed multiplet, where the isoscalar octet and singlet mix to form ${\overline n}n$ and ${\overline s}s$ states, that are close to the $\omega$ and $\phi$ we observe. But, of course, hadrons are not just ${\overline q}q$ systems. Their Foch space includes additional ${\overline q}q$ pairs that largely correlate into the mesons into which the hadron decays. This is typified by the Schwinger-Dyson equation for the  propagator of the $\rho$ shown in Fig.~2. At lowest order it is a pure ${\overline q}q$ state that does not decay, and gives a pole on the real axis seen in Fig.~3. As additional ${\overline q}q$ pairs are created~\cite{loops}, the $\rho$ can decay --- predominantly to $\pi\pi$. This moves its pole onto the nearby unphysical sheet as in Fig.~3. Because the coupling to $\pi\pi$ has a $P$-wave suppression, the $\rho$ remains largely in a ${\overline q}q$ configuration and only a few percent of the time is it a $\pi\pi$ system. Consequently, the underlying quark model state is easily recognised.

\begin{figure}[h]
\includegraphics[width=0.5\textwidth]{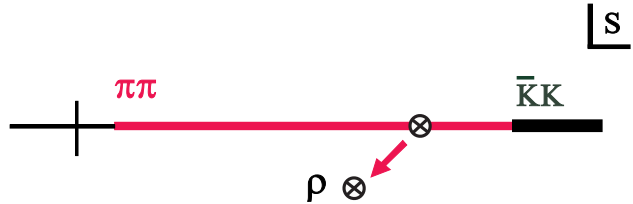}
\caption{The analytic structure of the $\rho$-propagator in the complex $s$-plane (where $s=$ momentum squared, corresponding to the graphs of Fig.~2). These have a cut at $\pi\pi$, $4\pi$, ${\overline K}K$, {\it etc.} thresholds, with the $\pi\pi$ channel being the most important. At lowest order, the propagator is real with a pole on the real axis corresponding to a bare ${\overline q}q$ meson. The corrections at higher orders, dominated by pion loops, give the full propagator with a pole on the nearby unphysical sheet. }
\end{figure}

\vspace{0.5mm}

In complete contrast, the $\sigma$ almost decays before it is born. It is 90\% a $\pi\pi$ system. Whether its underlying \lq\lq seed'' is ${\overline u}u+{\overline d}d$ or $[{\overline{ud}}][ud]$ or a glueball is very difficult to disentangle. Different modellings suggest different possibilities, but which is right is less important for the physics of the $\sigma$ than the dominance of its four quark component in a $\pi\pi$ configuration. Much the same can be said of the $f_0(980)$. It behaves almost everywhere as a ${\overline K}K$ system~\cite{weinstein}, as probably does the $a_0(980)$ (though in this case much less is known definitively).  

In a theorist's favourite world in which the number of colours, $N_c \gg 3$, ${\overline q}q$ and tetraquark states become quite distinct. As $N_c$ increases, the simple quark model state becomes narrower and more stable. The loop graphs in Fig.~2 are increasingly suppressed. In contrast,  a tetraquark state becomes wider, and more short lived, merging with the two meson continuum. How this applies to the scalars has been discussed by Jaffe~\cite{jaffe-4q}, and by Peleaz and collaborators~\cite{rios} with the most recent results presented at this meeting by Ruiz de Elvira~\cite{jacobo}.

Van Beveren, his collaborators and others~\cite{vanbev,tornqvist,japanese} 
have long highlighted how states with large couplings to decay channels
 can be dynamically generated. Thus a bare ${\overline q}q$ nonet up at 1.5 GeV,
 coupled through a system of equations like that in Fig.~2, can produce 
a nonet of hadrons close to this mass region and a second set of states 
dominated by their decay channels sitting much closer to 1 GeV --- see Fig.~4 as an illustration. These are arranged, not in the pattern of an ideal nonet that seeded them, but according to the hadronic channels that dominate their existence: $\pi\pi$ for the $\sigma$, $K\pi$ for the $\kappa$ and ${\overline K}K$ for the $f_0$ and $a_0$. These look very like the tetraquark pattern of Fig.~1, but are in fact seeded by the higher mass ${\overline q}q$ states
{\footnote{A comment is in order about the effect on the resonance propagator 
in Fig.~2 of closed hadronic channels. In principle, the infinity of such channels 
renormalises the resonance pole position by infinite amounts. The 
renormalised mass of the seed states then takes all these closed channels 
into account. Consequently, it is only the nearby open ones
which produce finite renormalizations that need be computed. 
This can be formally defined by renormalising the \lq\lq seed'' masses 
at the first strongly coupled threshold. So, for example, for the 
${\overline s}n$ and ${\overline n}s$ states their seed mass is 
defined at $K\pi$ threshold. Closed channels are then subtracted, 
as in ~\cite{wilson}.}.  

\vspace{3mm}
\begin{figure}[h]
\includegraphics[width=0.60\textwidth]{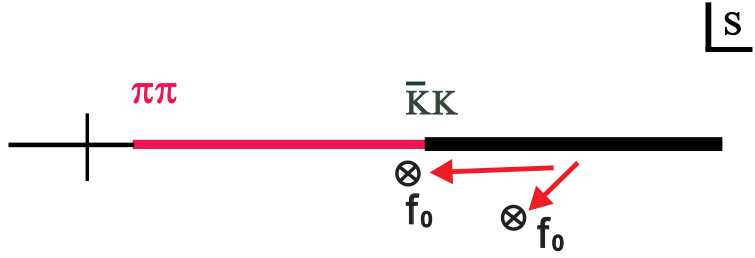}
\vspace{-2mm}
\caption{The analytic structure of the $f_0$-propagator in the complex $s$-plane (where $s=$ momentum squared, corresponding to the graphs of Fig.~2). These have a cut at $\pi\pi$, $4\pi$, ${\overline K}K$, {\it etc.} thresholds. If the $f_0$ \lq\lq seed'' is ${\overline{s}}s$, then the ${\overline{K}}K$ channel is the most important. The corrections, dominated by kaon loops, give the full propagator with poles on the nearby unphysical sheet. In the calculations of~\cite{vanbev} one is close to ${\overline{K}}K$ threshold. }
\end{figure}

\section{Scalars in semileptonic decays of heavy flavours}

While  long distance probing of the light scalars sees their {\it hadron molecule} nature~\cite{weinstein}, shorter distance interactions might reveal their intrinsic seeds.
Semileptonic decays, especially of heavy flavours, appear to provide just such a probe. In $B$ or $D$ decay, the heavy quark changes into a light quark by the emission of a $W$ that materialises as a lepton pair, as in Fig.~5. This interaction takes place over a distance scale of $\sim 0.01$ fermi. $D$ and $D_s$ decays (Fig.~5) produce scalars~\cite{stone} that decay to $\pi\pi$ or ${\overline K}K$. From these decay patterns, one can deduce the composition of the scalar resonances $f_0$ if they are ${\overline q}q$. Applying this to the CLEO-c results, Ecklund {\it et al.}~\cite{ecklund} find the $f_0(980)$ is largely ${\overline s}s$. 
\vspace{4mm}
\begin{figure}[h]
\includegraphics[width=0.63\textwidth]{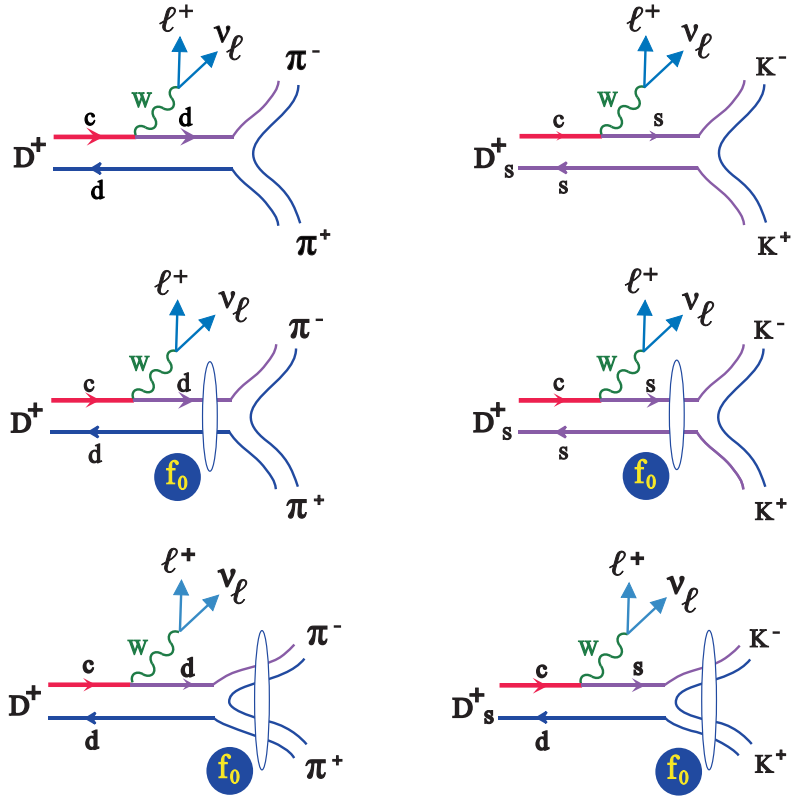}
\caption{Quark line representation of the semileptonic production of $D^+$ and $D_s^+$ to ${\overline d}d$ and ${\overline s}s$ systems, respectively, decaying to $\pi\pi$ and ${\overline K}K$ final states. The graphs in lines 2 and 3 illustrate $f_0$ bound states being formed in either ${\overline q}q$ or ${\overline{qq}}qq$ configurations.}
\end{figure}

Since tetraquark states have a different mixing pattern, as shown in Fig.~1, Wang and Lu~\cite{wang-lu} have proposed that the same idea can be used to judge whether the $f_0$ states are two quark or four quark states (see the lower 4 graphs in Fig.~5). While the short distance nature of the weak decay appears to probe the \lq\lq primordial seeds'', this basic process is dressed over longer timescales by gluon and quark interactions, Fig.~6. Whether this mechanism  can distinguish a two quark seed from a four quark one is then a mute point. The additional ${\overline q}q$ pair required for a tetraquark meson is not necessarily created soon after the weak interaction. If later, then the \lq\lq model-independent''  distinction proposed in ~\cite{wang-lu}  between a tetraquark state and a two meson final state is inevitably lost. Indeed for the lightest scalars (Fig.~1), one would expect the decay pattern to be that of {\it hadronic molecules} moderated only by the difference in available phase-space.

\begin{figure}[h]
  \includegraphics[width=0.76\textwidth]{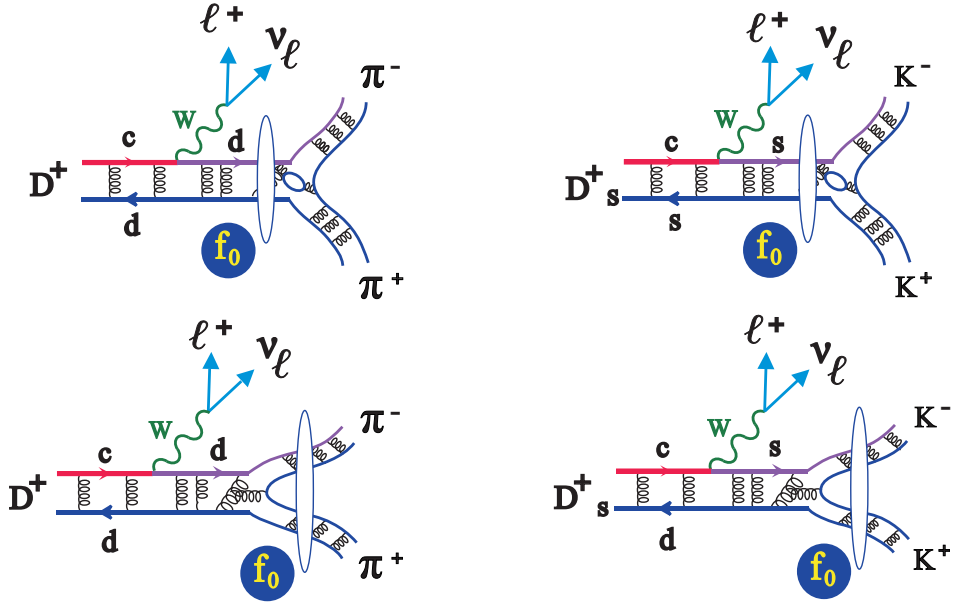}
\caption{Quark line representation of the semileptonic production of $D^+$ and $D_s^+$ to ${\overline{d}}d$ and ${\overline{s}}s$ components of a resonant $f_0$, or with additional quark pairs a tetraquark $f_0$ meson. The basic weak interaction is dressed by the longer range gluon interactions required to produce hadronic binding. In each case the $f_0$ decays to $\pi\pi$ and ${\overline{K}}K$ final states.}
\end{figure}
\vspace{2mm}

Nevertheless, semileptonic decays are able to provide information on meson-meson reactions we cannot access in other ways. Not only do they allow near threshold $\pi\pi$ interactions to be studied in $K_{e4}$ decays, but hold out the prospect of insight into $K\pi$ interactions too.
Most of our present information about such interactions come from high energy $Kp$ collisions that produce a $K\pi$ system at the very small momentum transfers dominated by one pion exchange. The highest statistics experiment from LASS provides almost all we know about $K^-\pi^+\to K^-\pi^+$ scattering from 200 MeV above threshold to 2 GeV. The cross-section shows the expected peaks from the $K^*(890)$ with spin-1 and the tensor $K_2(1430)$. When combined with results on $K^+\pi^+$ production also taken at SLAC, one can extract the pure $I=1/2$ signal. The LASS partial wave analysis~\cite{lass} established the broad scalar $K^*_0(1430)$ with a steadily rising phase from 825 MeV, but no sign of any narrow $\kappa(900)$ as confirmed in ~\cite{cherry}.

Extending chiral perturbation theory from 2 to 3 flavours allows these data to be continued to threshold and to the nearby crossed channel cut. This has allowed Descotes-Genon and Moussallam~\cite{descotes} to locate a $\kappa$ pole very close to $K\pi$ threshold at $E\,=\,658 -i 289$ MeV. Like the $\sigma$ this state decays very fast. In principle, the
semileptonic decay $\,D^\pm\to \ell^\pm \nu_{\ell} (K \pi)\,$ can confirm these results and even add new information, checking the extension of chiral dynamics to the heavier strange quark. The  hadronic final state interactions observed in these decays teach us about the phases of $K\pi$ scattering in the region of elastic unitarity. The FOCUS experiment at Fermilab investigated this with $\ell =\mu$ in the $K\pi$ mass region from 800 to 1000 MeV around the $K^*(890)$. FOCUS~\cite{focusdl4} showed that their observed forward-backward asymmetry  is consistent with an $S$-wave having a phase of $\sim 45^o$ around 900 MeV, exactly as found by LASS. Their limited statistics do not allow a detailed analysis outside the $K^*$ region.
However, BaBar (with $\ell = e$) promises sufficient events to determine the $S-P$ phase difference from $K\pi$ threshold to 1.6 GeV. Results should be reported shortly~\cite{patrick}. Though very exciting, these will still lack the precision needed to impact on the determination of the $\kappa$-pole. Hadronic decays however will.  

\section{Scalars in the hadronic decays of heavy flavours}

Semileptonic decays have the beauty of just two body hadronic final state interactions as in Figs.~5,6. However, involving neutrinos,  they may never achieve the precision available in purely hadronic decays. Processes like $D\to K\pi\pi$ now have tens if not hundreds of thousands of fully reconstructed events. The remarkable thing about these decay distributions, which are usefully displayed in a Dalitz plot, like that in Fig.~7, is that they are not uniform. The $D$ does not decay to $K\pi\pi$ directly. Rather the Dalitz plot typically shows clear bands in $K\pi$ and (in the case of $D^0$ decay) in $\pi\pi$ masses too. This indicates that the decay is dominated by two body processes, both $D \to \pi K^*(890) \to  \pi\pi K$ and $D \to K \rho \to K\pi\pi$, as in Fig.~8. While the  $K^*$ and $\rho$ are the most obvious (and narrow) isobars, the distribution is in fact controlled by broader $0^{++}$ states, both strange and non-strange. Indeed, it is typical of almost all heavy flavour decays that they are dominated by scalars, and an accurate description of these is essential for extracting precise information about the relevant CKM matrix element. 
\begin{figure}[h]
\includegraphics[width=0.70\textwidth]{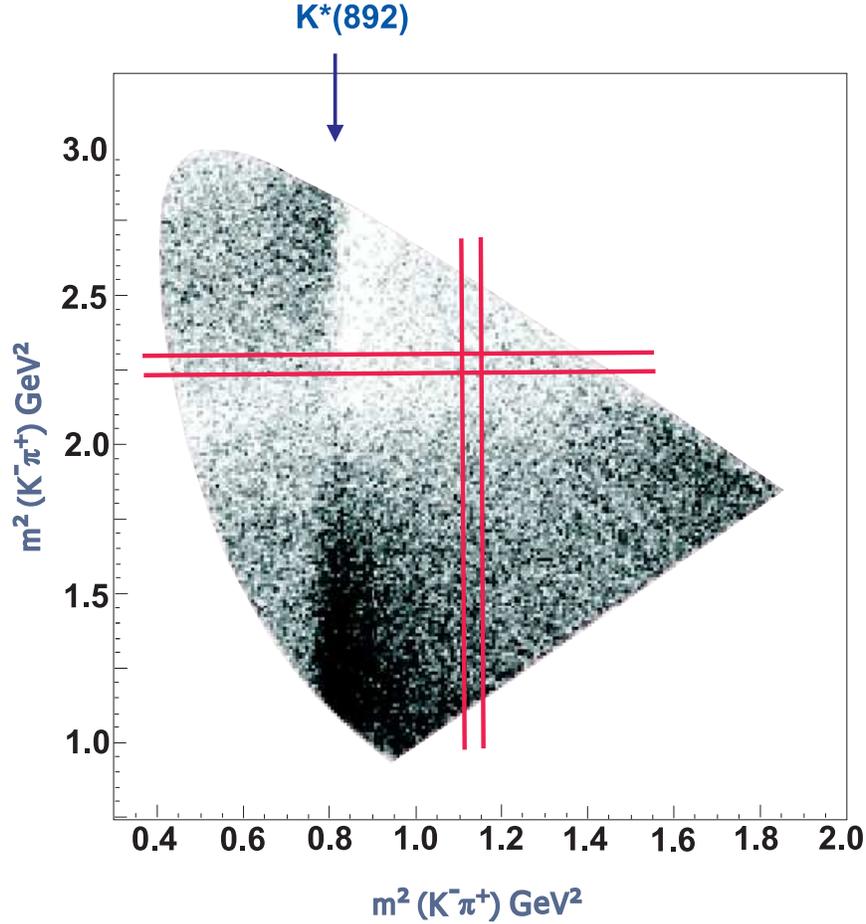}
\caption{Dalitz plot for $D^+\to K^-\pi^+\pi^+$ from FOCUS. Since it is symmetric under the interchange of the two pions, only half the full plot is shown. The plot is divided into bands of fixed $K\pi$ mass for a model-independent partial wave analysis, like that of ~\cite{e791brian,e791pub,bediaga}.   }
\end{figure}

$CP$ violation in the $B$ system can be studied by comparing $B\to D {\overline K}$ with $B \to {\overline D}K$, where the $D \to K\pi\pi$, when in each case one has a common
${\overline K}K\pi\pi$ final state. The key to a precision study is an accurate understanding of the $D$ decay~\cite{belle1,babar1}, which is dominated by the scalars. The scalars being broad and overlapping are not described by sums of simple Breit-Wigners. 
Indeed, such are the current statistics that, even in the $P$-wave $K\pi$ system, one has to know how to add the contribution
 of the $K^*(890)$ and the $K^*_1(1430)$ to ensure a sufficiently precise description of their overlap in the 1-1.1 GeV region.
 The amplitudes in every wave in the decay are better represented by a $P$-vector with the two body final state interactions described by a $K$-matrix formalism that adequately represents all we know of the same interactions in other production processes.

\begin{figure}[h]
\includegraphics[width=0.77\textwidth]{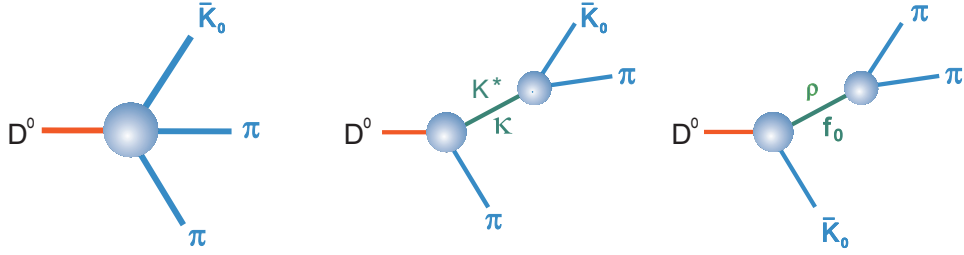}
\caption{Three body decay of a parent particle here $D^0$ to ${\overline{K^0}}\pi^+\pi^-$, and its representation in terms of isobars, in which intermediate $K\pi$ and $\pi\pi$ resonances are formed, which then subsequently decay.}
\vspace{-0.3cm}
\end{figure}

Progress in learning about the $S$-wave $K\pi$ interaction can be made by studying a decay like $D^+\to K^-\pi^+\pi^+$, Fig.~7. Rather than fitting with an unsatisfactory model for the $S$-wave~\cite{e791kappa}, E791~\cite{e791brian,e791pub} (and more recently FOCUS~\cite{bediaga}) have parametrised the $K\pi$ $S$-wave by a magnitude and phase in slices across the Dalitz plot as in Fig.~7.
 The interference between these bands and between one $S$-wave final state in one $K\pi$ channel with modelled $P$ and $D$-waves  in the other channel determines the $S$-wave magnitude and phase. In Fig.~9 we show the results for the $S$-wave phase as found by FOCUS~\cite{bediaga}, which agree closely with those found with lower statistics by E791~\cite{e791pub}. In these analyses this phase is determined relative to that of the $P$-wave fixed to be $90^o$ at 892 MeV. In making the plot shown here we have shifted these phases up by $100^o$ to make the phase effectively zero at $K\pi$ threshold. In Fig.~9 we compare this with the $I=1/2$ $S$-wave phase found from the LASS experiment on scattering above 825 MeV and continued to threshold according to one loop chiral perturbation theory. One sees that though they have a common trend, these phases are different. This difference can come from several sources: (i) the $K^-\pi^+$ interaction in $D$-decay need not be pure $I=1/2$, and (ii) there can be significant rescattering contributions.

\vspace{4mm}
\begin{figure}[h]
\vspace{5mm}
\includegraphics[width=0.55\textwidth]{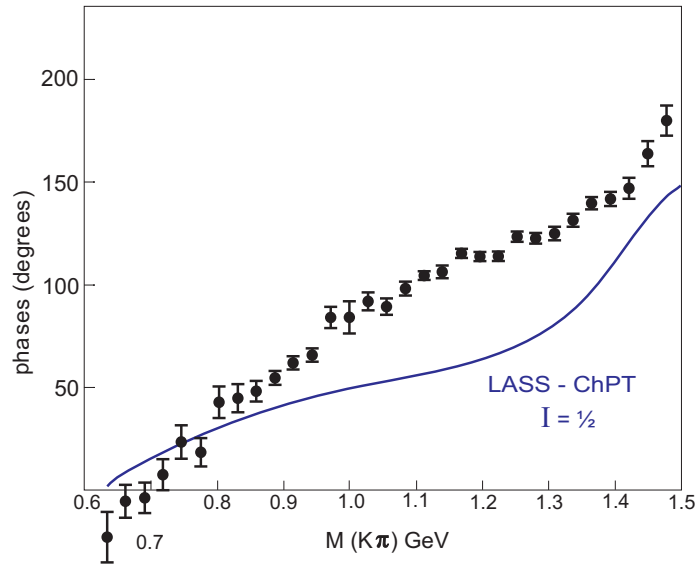}
\caption{ The $S$-wave $K\pi$ phase as a function of $K\pi$ mass. The solid curve represents a fit to the LASS results consistent with Chiral Perturbation Theory~\protect\cite{descotes} for the $I=1/2$ $S$-wave $K\pi$ phase from~\cite{malvezzi}. This is compared
with results of the Model Independent Partial Wave Analysis of FOCUS data by Link {\it et al.}~\cite{bediaga} shifted by $+100^o$. }
\end{figure}
\vspace{3mm}

For (i) the $D$-decay can have an $I=3/2$ component. While the relative contribution  of $I=1/2:3/2$  is fixed in $K^-\pi^+$ elastic scattering simply by 
Clebsch-Gordan coefficients, that is not the case in $D$-decay and is only fixed by additional information, as discussed in ~\cite{edera}. The second \lq\lq rescattering'' component occurs when the $D$ emits a $\pi$ forming a $K\pi$ system, which having interacted separates, and  the final state $K$ and $\pi$ rescatter on the spectator $\pi$, as indicated on the right hand side of Fig.~10. Two body unitarity imposes the constraint shown in Fig.~10, in which all  three graphs contribute to the imaginary part of the decay amplitude in each partial wave.

\begin{figure}[h]
\includegraphics[width=0.92\textwidth]{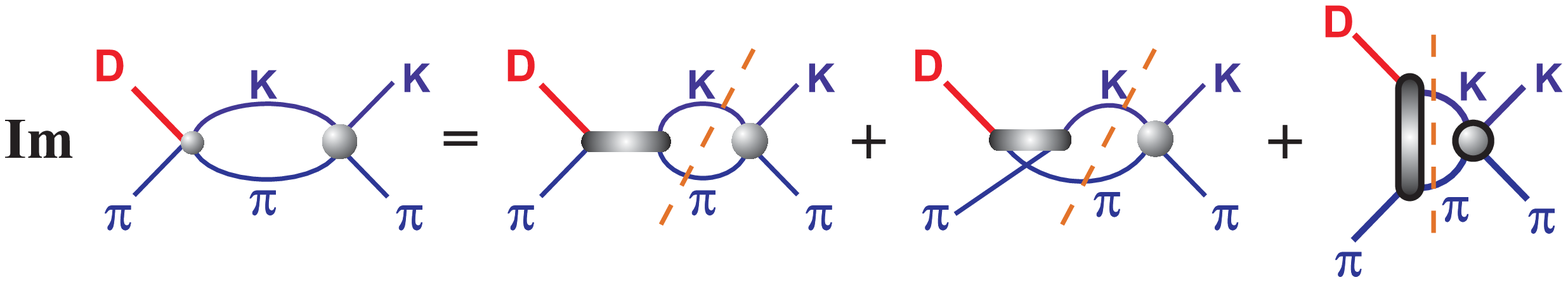}
\caption{Unitarity for the $K\pi$ system in $D$-decay in the elastic region.
The dashed lines denote the particles in the intermediate states are on mass shell.}
\end{figure}

If there were no rescattering, then only the first term on the right of Fig.~10 would contribute and the phase of the $K\pi$ interaction 
amplitude in each spin and isospin
 in both $D$ decay and $K\pi$ scattering would be equal in the region of elastic unitarity. This is effectively up to $K\eta'$ threshold. The difference seen in Fig.~9 might indicate rescattering effects from the other two graphs in Fig.~10 should not be neglected.  A major task is then how to build the unitarity constraint of Fig.~10 into the Dalitz analysis, so we can be sure we are treating the basic meson-meson final state interactions correctly. This is essential if we are going to learn about \lq\lq new'' physics in precision experiments whether in $J/\psi$-decays at BESIII,  in ${\overline p}p$ collisions at FAIR, in the photoproduction of multi-meson final states with {\it GlueX} at JLab and in $CP$ violation studies at {\it LHCb}. A worldwide effort to develop the necessary analysis tools is now taking shape. This is essential if these forthcoming experiments are to deliver their objectives of revealing novel physics.

\section{Low mass scalars and their two meson components}

The light scalars, nineteen in number (two isotriplets, four isodoublets and 5 isosinglets~\cite{pdg}) might be thought to form two nonets with a glueball left over, as displayed in Fig.~1. It has become popular to think of these as a ${\overline q}q$ multiplet around 1.4 GeV and a ${\overline{qq}}qq$ nonet below 1 GeV. This qualitatively explains the observed flavour structure. Tests of the expected different mixing schemes for the upper and lower multiplets have been proposed. However, we have argued here that the lower states
actually spend most of their time in the di-meson configurations, which dominate their decays. They are rather rarely to be observed in their ``primordial''  ${\overline q}q$ or $\overline{qq}qq$ or gluonic states.

A way of studying the charged  constituents of a hadron is to measure its coupling to photons. Scalars, like tensors, couple to two photons. Their radiative width measures the square of the average charge squared of their constituents. This works perfectly for the $f_2(1270)$, $a_2(1320)$ and $f_2'(1525)$ that form an almost ideally mixed tensor multiplet.  $B$-factories, in particular Belle~\cite{Bellegam}, have produced two photon results of unprecedented precision. These have the power to determine the radiative width of light states up to 2 GeV  that decay to $\pi\pi$, $\pi\eta$ and ${\overline K}K$ --- with the planned upgrade to the KLOE detector at the DA$\Phi$NE machine at Frascati adding yet further precision below 1.2 GeV~\cite{kloe2}.

\vspace{6mm}
\begin{figure}[h]
\includegraphics[width=0.29\textwidth]{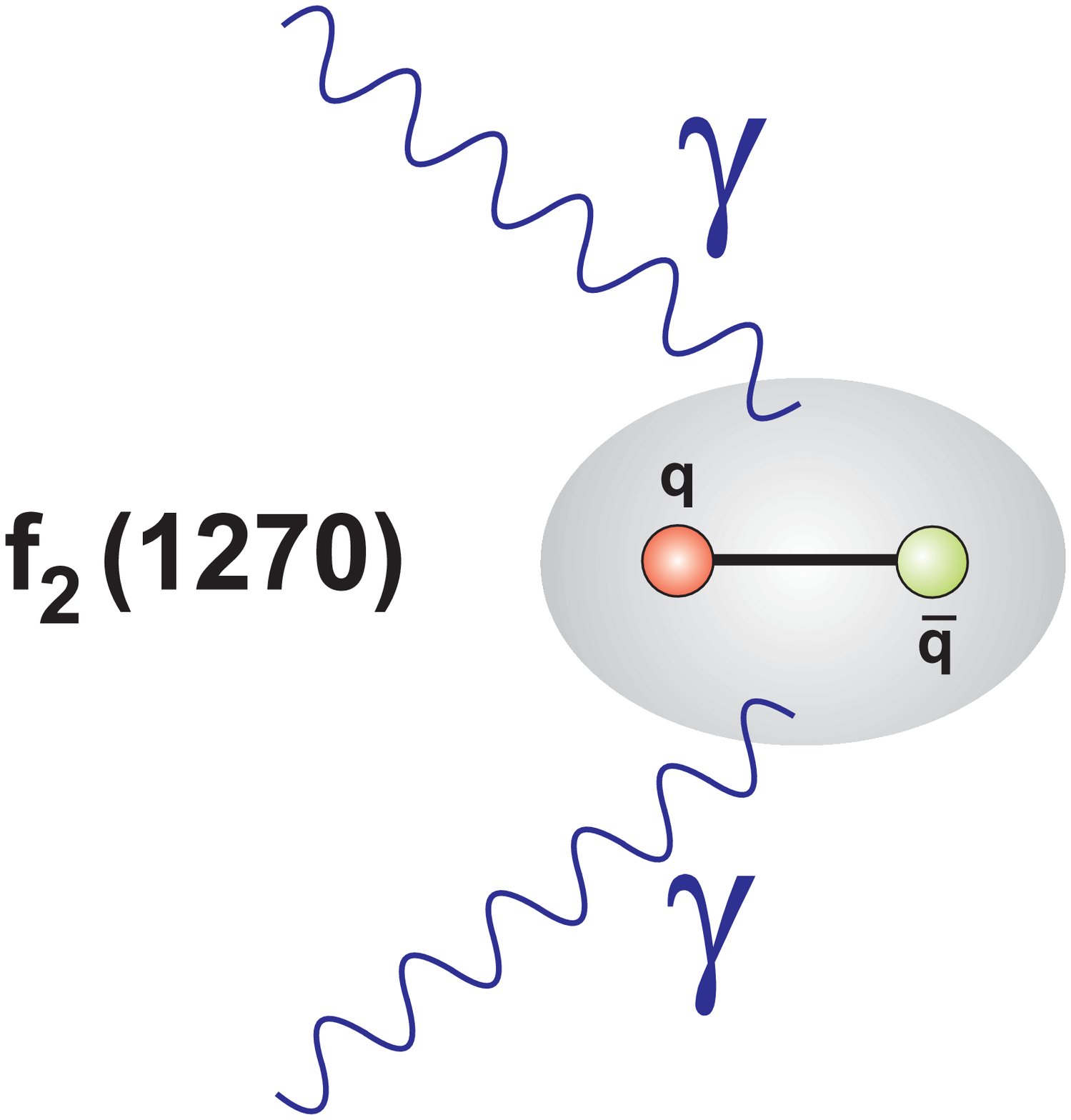}\hspace{20mm}
\includegraphics[width=0.33\textwidth]{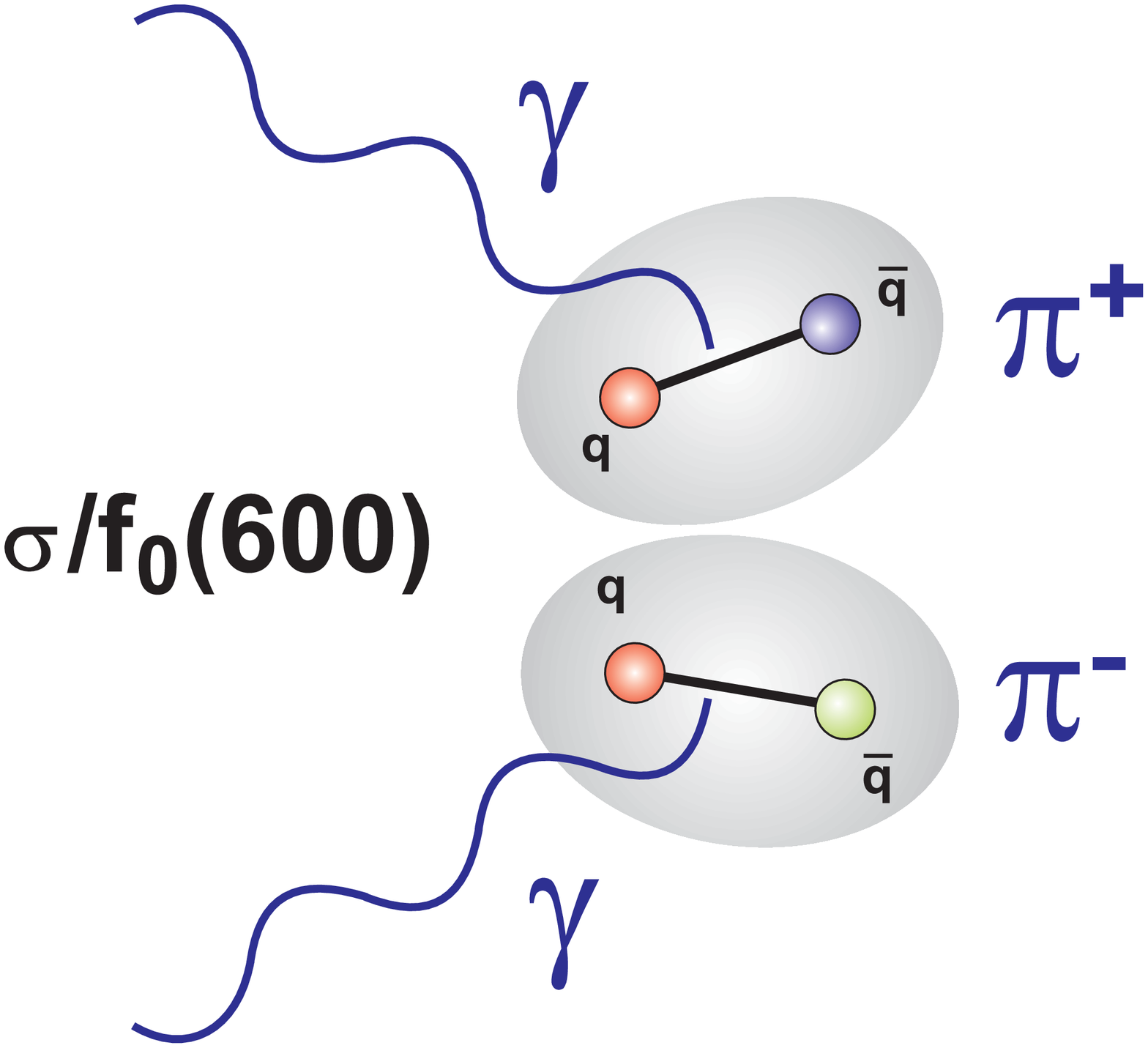}
\caption{Illustration of two photons coupling to a largely ${\overline q}q$  meson like the $f_2(1270)$ and a state dominated by its two meson decay like the $\sigma/f_0(600)$. In the latter case, the photons coupling to the decay final state 
 dominates over any coupling to the {\it intrinsic} make-up of the state, regardless of its composition.}
\end{figure}
\vspace{4mm}

 In a two photon collision, the photons at 1.3 GeV, where the tensors lie, have short enough wavelength to probe the constituent quarks, Fig.~11. However, at lower energies, particularly in the $\sigma$ region, the photons most often see the pions to which this scalar decays, Fig.~11.
Calculation combining analyticity, unitarity, crossing symmetry and the low energy theorem of QED shows that the radiative width of the $\sigma$ is indeed dominated by its two pion contribution~\cite{mrp-prl,oller}. Much the same happens for the $f_0(980)$ which is controlled by its large ${\overline K}K$ component. Two photon interactions, as in many other experiments, see the whole Foch space at once, and the seed (in the sense of the first term on the right of the propagator equation of Fig.~3) is not readily separated. Removing the two meson loop component might be thought to probe this \lq\lq intrinsic'' contribution~\cite{mink}. Only models can determine this component. Mennessier, Narison and Ochs~\cite{mink}, for instance, claim that its two photon coupling being small points to a gluonic seed. 

Experiment reveals
the light scalars to be largely two meson states. They may be seeded by ${\overline q}q$, ${\overline{qq}}qq$ or a glueball. While a tetraquark nonet might seem appealing. The states are largely in two meson form. The work of van Beveren and friends~\cite{vanbev} suggests that the quark model seeds, that underlie the higher mass scalars, dynamically generate the two meson states too. Being dynamically generated they reside close to the thresholds of the channels to which they most strongly couple. One might think that an effect of the seeds is felt in the \lq\lq counting'' of states. However, whether ${\overline q}q$ and glueball seeds all generate lower mass scalars is a question of dynamics (for an imperfect illustration see~\cite{boglione}), and so mere counting may not be sufficient. More work is needed.

Precision description of these states is essential for not just understanding their nature but unravelling their role in translating the dynamics of quarks and gluons to the hadron world. A role that very many experiments in the next 5-10 years
at BESIII, LHCb, GlueX and FAIR will illuminate with unprecedented statistics. Analysis techniques to match these experimental advances will be essential if we are to find out more about these elusive Scalargators and reveal the intriguing physics they embody themselves and by their dominance the physics beyond the Standard Model they  overshadow.


\begin{theacknowledgments}
It is a pleasure to thank the organisers for the invitation to return to 
Tallahassee almost 37 years after I gave my first conference talk there.
David Morgan chaired the session in which I spoke. He later became my long time collaborator. David, from whom I learnt such a great deal, sadly died earlier this year. This talk is dedicated to him, for his wisdom, understanding and friendship. 

I acknowledge partial support of the EU-RTN Programme, Contract No. MRTN--CT-2006-035482, \lq\lq Flavianet'' for this work.

\end{theacknowledgments}



\bibliographystyle{aipproc}   

\end{document}